\documentclass[fleqn,a4paper,12pt]{article}
\usepackage{amsmath}
\usepackage{amssymb}
\usepackage{mathtools}
\usepackage{bm}
\usepackage{amsthm}
\usepackage[linesnumbered,ruled,vlined]{algorithm2e} 
\usepackage{graphicx}
\usepackage[round,sectionbib]{natbib}
\usepackage{setspace} 
\usepackage{a4wide}
\usepackage{url}
\input{ee.sty}

\newtheorem{prp}{Proposition}
\newtheorem{lem}{Lemma}

\DeclareMathOperator*{\argmin}{arg\,min}
\DeclareMathOperator*{\argmax}{arg\,max}
\DeclarePairedDelimiter{\floor}{\lfloor}{\rfloor}
\allowdisplaybreaks
\title{
	\vspace{0in}
	\textmd{\fontsize{40pt}{11pt}{\textbf{Sparse Unit-Sum Regression}}}\\
	\vspace{0in}
}

\date{October 9th, 2018}
\author{N.W. Koning\footnote{University of Groningen, n.w.koning@rug.nl. Address: Nettelbosje 2, 9747 AE Groningen, Netherlands. Phone: +31503639106},\quad P.A. Bekker\footnote{University of Groningen}}

\begin{document}
	\maketitle

	\begin{abstract}
		\noindent This paper considers sparsity in linear regression under the restriction that the regression weights sum to one. We propose an approach that combines $\ell_0$- and $\ell_1$-regularization. We compute its solution by adapting a recent methodological innovation made by \citet{bertsimas2016best} for $\ell_0$-regularization in standard linear regression. In a simulation experiment we compare our approach to $\ell_0$-regularization and $\ell_1$-regularization and find that it performs favorably in terms of predictive performance and sparsity. In an application to index tracking we show that our approach can obtain substantially sparser portfolios compared to $\ell_1$-regularization while maintaining a similar tracking performance. \\

	\end{abstract}
		
	\section{Introduction}
	Linear regression with coefficients that sum to one (henceforth \textit{unit-sum regression}) is used in portfolio optimization and other economic applications such as forecast combinations \citep{timmermann2006forecast} and synthetic control \citep{abadie2010synthetic}. \\
	
	In this paper, we focus on obtaining a sparse solution (i.e. containing few non-zero elements) to the unit-sum regression problem. A sparse solution may be desirable for a variety of reasons, such as making a model more interpretable, improving estimation efficiency if the underlying parameter vector is known to be sparse, remedying identification issues if the number of variables exceeds the number of observations, or application-specific reasons such as reducing cost by limiting the amount of constituents in a portfolio. \\
	
	A popular method to produce sparsity is to use regularization. Theoretically, the most straightforward way to obtain a sparse solution is to use $\ell_0$-regularization (also known as best-subset selection), which amounts to restricting the number of non-zero elements in the solution. However, the use of $\ell_0$-regularization is NP-hard \citep{coleman2006minimizing, natarajan1995sparse} and has traditionally been seen as computationally infeasible for problems with more than about 40 variables, both in unit-sum regression and standard linear regression. \\
	
	Due to these computational difficulties, $\ell_0$-regularization has often been replaced by $\ell_1$-regularization, also known as Lasso \citep{tibshirani1996regression}. In $\ell_1$-regularization, the $\ell_0$-norm restriction that restricts the number of non-zero elements is replaced by an $\ell_1$-norm restriction that restricts the absolute size of the coefficients. This turns the problem into an easier to solve convex optimization problem. An $\ell_1$-norm restriction shrinks the weights towards zero and, as a consequence of the shrinkage, produces sparsity by setting some weights exactly equal to zero. \\

	The use of $\ell_1$-regularization in the presence of a unit-sum restriction was first considered by \citet{demiguel2009generalized} and \citet{brodie2009sparse} in the context of portfolio optimization. They show that $\ell_1$-regularization is able to produce sparsity in combination with a unit-sum restriction. In addition, they demonstrate that the combination can be viewed as a restriction on the sum of the negative weights. In some applications it is highly desirable to have a parameter that explicitly controls the sum of the negative weights. For example, in a portfolio optimization context negative weights represent potentially costly short positions. \\
	
	However, the unit-sum restriction causes a problem when using $\ell_1$-regularization: due to the unit-sum restriction the $\ell_1$-norm of the weights cannot be smaller than 1. This imposes a lower bound on the amount of shrinkage produced by $\ell_1$-regularization. In turn, this places an upper bound on the sparsity produced by $\ell_1$-regularization. This upper bound depends entirely on the data, which makes it difficult to rely on $\ell_1$-regularization if a specific level of sparsity is desired. In addition, due to the bound there does not always exist a value of the tuning parameter that guarantees the existence of a unique solution. Furthermore, \citet{fastrich2014cardinality} point out that a combination of a non-negativity restriction and a unit-sum restriction fixes the $\ell_1$-norm of the weights to 1, which renders $\ell_1$-regularization useless. \\
	
	In order to address these issues and obtain sparse solutions in unit-sum regression, we use a recent innovation in $\ell_0$-regularization in the standard linear regression setting by \citet{bertsimas2016best}. They show that modern Mixed-Integer Optimization (MIO) solvers can find a provably optimal solution to $\ell_0$-regularized regression for problems of practical size. To achieve this, the solver is provided with a good initial solution obtained from a discrete first-order (DFO) algorithm. In a simulation study, they show that $\ell_0$-regularization performs favorably compared to $\ell_1$-regularization in terms of predictive performance and sparsity. \\
	
	An extended simulation study comparing $\ell_0$- and $\ell_1$-regularization in the standard linear regression setting is performed by \citet{hastie2017extended}. They find that find that $\ell_0$-regularization outperforms $\ell_1$-regularization if the signal-to-noise ratio (SNR) is high, while $\ell_1$-regularization performs better if the SNR is low. Additionally, they find that if the tuning parameters are selected to optimize predictive performance, $\ell_0$-regularization yields substantially sparser solutions. \\
	
	A combination of $\ell_0$- and $\ell_1$-regularization ($\ell_0\ell_1$-regularization) is studied in the standard linear regression context by \citet{mazumder2017subset}. They observe that this combination yields a predictive performance similar to $\ell_1$-regularization if the SNR is low, and a predictive performance similar to $\ell_0$-regularization if the SNR is high. In addition, they find that $\ell_0\ell_1$-regularization produces more sparsity compared to $\ell_1$-regularization, if the tuning parameters are selected in order to optimize predictive performance. \\
	
	Motivated by the results in the standard linear regression setting, we propose the use of $\ell_0\ell_1$-regularization in unit-sum regression. Specifically, let $\vy$ be a $t$-vector and let $\mX$ be a $t \times m$ matrix, then we consider the problem
	\begin{align}
		\label{problem}
		\min_{\vbeta}\|\vy - \mX\vbeta\|_2^2,
		\quad\text{s.t. } \sum_{i=1}^{m} \beta_i = 1,
		\quad\|\vbeta\|_0 \leq k,
		\quad\|\vbeta\|_1 \leq 1 + 2s,
	\end{align}
	where $\beta_i$ are the elements of $\vbeta$, $\|\vbeta\|_0 = \sum_{i=1}^{m}1_{\{\beta_i \neq 0\}}$ is the $\ell_0$-norm of $\vbeta$, $\|\vbeta\|_1 = \sum_{i=1}^{m}|\beta_i|$ is the $\ell_1$-norm of $\vbeta$, $s \geq 0$ and $1 \leq k \leq m$. Notice that this problem is equivalent to $\ell_0$-regularized unit-sum regression if $s$ is sufficiently large, and equivalent to $\ell_1$-regularized unit-sum regression if $k = m$. \\
	
	The formulation in \eqref{problem} provides users with explicit control over both the sparsity of the solution and the sum of the negative weights of the solution. In addition, if the tuning parameters are selected in order to maximize predictive performance, we find in a simulation experiment that $\ell_0\ell_1$-regularization:
	\begin{itemize}
		\item performs better than $\ell_0$-regularization in terms of predictive performance, especially if the signal-to-noise ratio is low.  
		\item performs well compared to $\ell_1$-regularization in terms of predictive performance, especially for higher signal-to-noise ratios, while at the same time producing much sparser solutions.
	\end{itemize}
	
	The main contributions of this paper can be summarized as follows. [1] We propose $\ell_0\ell_1$-regularization for the unit-sum regression problem. [2] We analyze the problem for orthogonal design matrices and provide an algorithm to compute its solution. [3] We show how the algorithm for the orthogonal design case can be used in finding a solution to the general problem by extending the framework of \citet{bertsimas2016best} to unit-sum regression. [4] We perform a simulation experiment which shows that our approach performs favorably compared to $\ell_0$-regularization or $\ell_1$-regularization. [5] We demonstrate in an application to stock index tracking that a $\ell_0\ell_1$-regularization is able to find substantially sparser portfolios than $\ell_1$-regularization, while maintaining a similar out-of-sample tracking error. \\		
	
	The remainder of the paper is structured as follows. In Section \ref{compH}, problem \eqref{problem} is studied under the assumption that $\mX$ is orthogonal and an algorithm for the orthogonal case is presented. Section \ref{sec:sparsity} analyzes the sparsity production for the orthogonal case and yields some intuitions about the problem. Section \ref{sec:DFO} links the algorithm for the orthogonal case to the framework of \citet{bertsimas2016best} in order to find a solution to the general problem. In Section \ref{sec:num}, the simulation experiments are presented. Section \ref{subsec:maxsparse} provides an application to index tracking.
	
	\section{Orthogonal Design}\label{compH}
		As problem \eqref{problem} is difficult to study in its full generality, we first consider the special case that $\mX$ is orthogonal. We derive properties of a solution to \eqref{problem} under orthogonality and use these properties in order to construct an algorithm that finds a solution. The algorithm is presented at the end of the section. In Section \ref{sec:DFO} this algorithm is used in finding a solution to the general problem by extending the framework of \citet{bertsimas2016best}. In Section \ref{sec:sparsity} we analyze the sparsity of the solution under orthogonality.\\
		
		Assume that $\mX'\mX = \mX\mX' = \mI_m$, where $\mI_m$ is the $m \times m$ identity matrix. Let us write $\veta = \mX'\vy$, so that minimizing $\|\vy - \mX\vbeta\|_2^2$ in $\vbeta$ is equivalent to minimizing $\|\mX'\vy - \vbeta\|_2^2 = \|\veta - \vbeta\|_2^2 =: Q(\vbeta)$. Define 
		\begin{align}
			\calT:=\left\{\vbeta\in\mathbb{R}^m\ \left|\ \sum_{i=1}^m \beta_i=1,\ \|\vbeta\|_0\leqslant k,\ \|\vbeta\|_1 \leqslant 1 + 2 s\right.\right\}. \label{def:T}
		\end{align} 
		 Then, problem \eqref{problem} can be written as $\min_{\vbeta \in \calT} Q(\vbeta)$. \\
		
		We assume the elements of $\veta$ are different and $k < m$. Without further loss of generality we assume $\eta_1 > \eta_2 > \ldots > \eta_m$. In Section \ref{toegift} we relax the assumption that $k < m$ and allow for $k \leq m$. \\
		
		Let 
		\begin{align*}
			\calA_z:=\left\{\vbeta \in \mathbb{R}^m\ \left|\ \sum_{i=1}^m\beta_i=1,\ \|\vbeta \|_0\leqslant k,\ \|\vbeta \|_1=1+2z\right.\right\},
		\end{align*}
		where $0\leqslant z$, so that $\calT=\cup_{0\leqslant z\leqslant s}\calA_z$. If $\widehat{\vbeta}\in\argmin_{\vbeta \in \calT}Q(\vbeta)$, then $\widehat\vbeta \in \calA_z$ for some $z\leqslant s$. Let us denote this value of $z$ with $\hat{z}$. We will now show that $\widehat{\vbeta}$ can be computed from the signs of its elements and $\hat{z}$. In order to show this, we first solve a related problem and then show that $\widehat{\vbeta}$ is equal to the solution of a specific case of this related problem.  \\
		
		Let $\calP \subseteq \calM$ and $\calN \subset \calM$ be disjoint sets with cardinalities $p$ and $n$, respectively, where $\calM := \{1, \dots, m\}$. Define
		\begin{align*}
			\calB^{(\calP, \calN, z)} 
				:=\left\{\vbeta\in\mathbb{R}^m\ \left|\ \sum_{i\in\calP} \beta_i=1+z,\ \ \sum_{i\in\calN} \beta_i=-z,\ \mbox{and}\ \beta_i=0\  \mbox{if}\  i \in(\calP\cup\calN)^C\right.\right\}.
		\end{align*}
		Minimization of $Q(\vbeta)$ over the affinely restricted set $\calB^{(\calP, \calN, z)}$ has the solution 
		\begin{align}
		\vbeta^{(\calP,\calN,z)}:= 
			\argmin_{\vbeta\in\calB^{(\calP, \calN, z)}} Q(\vbeta) =
			\begin{cases}\label{cases}
				\eta_i - \frac{\left(\sum_{j\in\calP} \eta_j\right)-1-z}{p}, \qquad& i\in\calP,\\
				\eta_i - \frac{\left(\sum_{j\in\calN} \eta_j\right)+z}{n}, \qquad& i\in\calN,\\
				0, \qquad& i\in (\calP\cup\calN)^C.
			\end{cases}
		\end{align}
		Recall that $\widehat{\vbeta} \in \argmin_{\vbeta \in \calT} Q(\vbeta)$ and let $\widehat{\calP}:=\{i\ |\ \widehat \beta_i>0\}$ and $\widehat{\calN}:=\{i\ |\ \widehat \beta_i<0\}$. Furthermore, let $\calC$ be the set of vectors $\vbeta$ with elements that have the same signs as the elements of $\widehat\vbeta$, then $\calA_{\hat{z}}\cap\calC\subseteq\calB^{(\widehat{\calP}, \widehat{\calN}, \hat{z})}$. Notice that the difference between $\calA_{\hat{z}}\cap\calC$ and $\calB^{(\widehat{\calP}, \widehat{\calN}, \hat{z})}$ is that there are no sign restrictions on elements $\beta_i \in \calB^{(\widehat{\calP}, \widehat{\calN}, \hat{z})}$, for which $i \in (\calP \cup \calN)$. Consequently, 
		\begin{align*}
		Q(\widehat{\vbeta})
			= \min_{\vbeta\in\calA_{\hat{z}}\cap\calC}Q(\vbeta)
			\geqslant \min_{\vbeta\in\calB^{(\widehat{\calP}, \widehat{\calN}, \hat{z})}} Q(\vbeta)
			= Q(\vbeta^{(\widehat{\calP},\widehat{\calN}, \hat{z})}).
		\end{align*}
		However, if $\widehat{\vbeta}\neq\vbeta^{(\widehat{\calP},\widehat{\calN},\hat{z})}$, then $\vbeta(\phi):=\phi\vbeta^{(\widehat{\calP},\widehat{\calN},\hat{z})}+(1-\phi)\widehat{\vbeta}\in\calA_{\hat{z}} \cap \calC$ for sufficiently small $\phi > 0$. Furthermore, as $Q(\vbeta(\phi))$ is a parabola in $\phi$ with a minimum at $\lambda=1$, we find that $Q(\vbeta(\phi)) < Q(\widehat{\vbeta})$ for small $\phi > 0$. As $\widehat{\vbeta} \in \argmin_{\vbeta\in\calA_{\hat{z}}}Q(\vbeta)$, this is a contradiction. Hence, $\widehat{\vbeta}=\vbeta^{(\widehat{\calP},\widehat{\calN}, \hat{z})}$, which is our first result. 
		\begin{prp}\label{prp1}
			If $\widehat\vbeta\in\argmin_{\vbeta\in\calT}Q(\vbeta)$, then $\widehat\vbeta=\vbeta^{(\widehat{\calP}, \widehat{\calN}, \hat{z})}$. 
		\end{prp}
		So, the problem can be decomposed into finding the components of the triplet $(\calP, \calN, z)$ that minimizes $Q(\vbeta^{(\calP, \calN, z)})$. Next, we will study the properties of these components.
		
		\subsection{Properties of $Q(\vbeta^{(\calP, \calN, z)})$ as a function of $\calP$ and $\calN$}
			The sorting of $\veta$ reveals an ordered structure in the sets $\calP$ and $\calN$ that minimize $Q(\vbeta^{(\calP, \calN, z)})$. This structure is described in the following result.
			\begin{prp}\label{prp2}
				If $\vbeta^{(\calP, \calN, z)} \in \argmin_{\vbeta \in \calA_z} Q(\vbeta)$, then $\calP = \{1, \dots, p\}$ and $\calN = \{m-n+1, \dots, m\}$ if $n\geqslant 1$, and $\calN = \emptyset$ if $n = 0$. 
			\end{prp}
			The proof is given in the Appendix. For sets such as $\calP = \{1, \dots, p\}$ and $\calN = \{m-n+1, \dots, m\}$, we use the notation $\vbeta^{(p,n,z)}:=\vbeta^{(\calP,\calN,z)}$, as in \eqref{cases}. The following result shows that $p$ and $n$ should be maximized such that $\vbeta^{(p, n, z)} \in \calA_z$.
			
			\begin{lem}\label{lem1}
				If $\vbeta^{(\tilde{p}, \tilde{n}, z)} \in \calA_z$ and $\vbeta^{(p, n, z)} \in \calA_z$, where $\tilde{p} \leq p$, $\tilde{n} \leq n$, $\tilde{p} + \tilde{n} < p + n$, then $Q(\vbeta^{(p, n, z)}) < Q(\vbeta^{(\tilde{p}, \tilde{n}, z)})$.
			\end{lem}
			The proof is given in the Appendix. \\
			
			We will now consider the relationship between $z$ and the pair $(p, n)$. With reference to \eqref{cases}, let us consider the sets
			\begin{align}\label{Pz}
			\calP_z&:=\left\{q\ \left|\ \eta_q-\frac{\left(\sum_{i=1}^q\eta_i\right)-1-z}{q}>0\right.\right\},\\
			\calN_z&:=\left\{q\ \left|\ \eta_{m-q+1}-\frac{\left(\sum_{i=1}^q\eta_{m-i+1}\right)+z}{q}<0\right.\right\},\label{Nz}
			\end{align}
			with cardinalities $|\calP_z| = p_z$ and $|\calN_z| = n_z$. As
			\begin{align*}
			\eta_q-\frac{\left(\sum_{i=1}^q\eta_i\right)-1-z}{q}
			&=\frac{q-1}{q}\left(\eta_q-\frac{\left(\sum_{i=1}^{q-1}\eta_i\right)-1-z}{q-1}\right)\\
			&< \eta_{q-1}-\frac{\left(\sum_{i=1}^{q-1}\eta_i\right)-1-z}{q-1},
			\end{align*}
			we find $\calP_z = \{1, \dots, p_z\}$, and similarly $\calN_z = \{m - n + 1, \dots, m\}$ if $z > 0$ and $\calN_z = \emptyset$ if $z = 0$. Additionally, we find that $p_z$ is increasing in $z$, and similarly that $n_z$ is increasing in $z$. So, by Lemma \ref{lem1} we have following result for $p_z + n_z \leq k$.
			\begin{prp}\label{prp3}
				If  $\vbeta^{(\calP, \calN, z)} \in \argmin_{\vbeta \in \calA_z} Q(\vbeta)$ and $p_z+n_z\leq k$, then $\vbeta^{(\calP, \calN, z)} = \vbeta^{(p_z,n_z,z)}$.
			\end{prp}
			We will now analyze how $Q(\vbeta^{(p_z,n_z,z)})$ varies with $z$ if $p_z + n_z \leq k$, and use this to find a minimizer $\widehat{\vbeta} \in \argmin_{\vbeta \in \calT} Q(\vbeta)$ if $p_s + n_s \leq k$. The case that $p_s + n_s > k$ is treated separately in Section \ref{sssec:solutions}. 
		
		\subsection{Properties of $Q(\vbeta^{(p_z,n_z,z)})$ as a function of $z$ for $p_z + n_z \leq k$}\label{sssec:z}
			As $p_z$ and $n_z$ are integers, they increase discontinuously as $z$ increases. In this subsection we show that $Q(\vbeta^{(p_z, n_z, z)})$ and its derivative are continuous in $z$ despite these discontinuities in $p_z$ and $n_z$. This will allow us to show that $\vbeta^{(p_s, n_s, s)} \in \argmin_{\vbeta \in \calT} Q(\vbeta)$ if $p_s + n_s \leq k$. \\
			
			Let $z_1^+=-1$ and
			$z_{p}^+=z_{p-1}^++(p-1)(\eta_{p-1}-\eta_{p}) =\sum_{i=1}^{p-1}(\eta_i-\eta_p)-1$, for $p=2,\ldots,m$.
			We then find the ordering $z_1^+<z_2^+<\ldots< z_m^+$, and
			\begin{align}
			\eta_p&=\frac{\left(\sum_{i=1}^{p} \eta_i\right)-1-z_p^+}{p},\qquad\ \  p=1,\ldots,m, \nonumber\\
			\eta_{p+1}
			&=\frac{\left(\sum_{i=1}^{p}\eta_i\right)-1-z_{p+1}^+}{p}
			= \frac{\left(\sum_{i=1}^{p+1} \eta_i\right) - 1 - z_{p+1}^+}{p + 1}, \qquad p=1,\ldots,m-1.\label{ap+1}
			\end{align}
			Consequently, if $z_p^+<z\leqslant z_{p+1}^+$, then 
			\begin{align}\label{boundp}
				\eta_{p}&>\frac{(\sum_{i=1}^p \eta_i)-1-z}{p}\geqslant \eta_{p+1}.
			\end{align} 
			
			Similarly, let
			$z_m^-=0$ and
			$z_{m-n+1}^-=z_{m-(n-1)+1}^-+(n-1)(\eta_{m-n+1}-\eta_{m-(n-1)+1})=\sum_{i=1}^{n-1} (\eta_{m-n+1}-\eta_{m-i+1})$, for $n=2,\ldots,m$.
			Then $z_{m-1+1}^-<z_{m-2+1}^-<\ldots<z_{m-m+1}^-$ and
			\begin{align}
			\eta_{m-n+1} 
				&=\frac{\left(\sum_{i=1}^n \eta_{m-i+1}\right)+z_{m-n+1}^-}{n},\qquad n=1,\ldots,m,\nonumber\\
			\eta_{m-n}
				&=\frac{\left(\sum_{i=1}^n\eta_{m-i+1}\right)+z_{m-n}^-}{n}
				= \frac{\left(\sum_{i=1}^{n+1} \eta_{m-i+1}\right) + z_{m-n}^-}{n+1},\qquad n=1,\ldots,m-1.\label{an+1}
			\end{align}
			Consequently, if $z_{m-n+1}^-<z\leqslant z_{m-n}^-$, then 
			\begin{align}\label{boundn}
				\eta_{m-n+1}&<\frac{(\sum_{i=1}^n \eta_{m-i+1})+z}{n}\leqslant \eta_{m-n}.
			\end{align} 
			
			Using the cardinalities $p_z$ and $n_z$ of the sets $\calP_z$ and $\calN_z$ in \eqref{Pz} and \eqref{Nz}, let $z_m:=\min\{z\ |\ p_z+n_z=m\}$. If $0 \leqslant z < z_m$, then $z_{p_z}^+<z\leqslant z_{p_z+1}^+$. If $ 0< z < z_m$, then $z_{m-n_{z}+1}^-<z\leqslant z_{m-n_z}^-$. The loss function
			\begin{align*}
			&\hspace{-.75cm}Q\left(\vbeta^{(p_z,n_z,z)}\right)=p_z\left\{\frac{\left(\sum_{i=1}^{p_z} \eta_i\right)-1-z}{p_z}\right\}^2+I_{\{n_z\geqslant 1\}}n_z\left\{\frac{(\sum_{i=1}^{n_z} \eta_{m-i+1})+z}{n_z}\right\}^2+\sum_{i=p_z+1}^{m-n_z}\eta_i^2
			\end{align*}
			is a continuous function of $z$ for $0 \leqslant z \leqslant z_m$, with derivative
			\begin{align} \label{der0}
			\frac{\rd Q\left(\vbeta^{(p_z,n_z,z)}\right)}{\rd z}&=-2\left\{\frac{\left(\sum_{i=1}^{p_z} \eta_i\right)-1-z}{p_z}\right\}+2I_{\{n_z\geqslant 1\}}\left\{\frac{(\sum_{i=1}^{n_z} 
			\eta_{m-i+1})+z}{n_z}\right\},
			\end{align}
			which is continuous for $0<z<z_m$. That is, using \eqref{ap+1}, if $z \uparrow z_{p_z+1}^+$, then 
			\begin{align*}
			-\frac{\left(\sum_{i=1}^{p_z} \eta_i\right) - 1 - z}{p_z} \uparrow \eta_{p_z+1}
			\end{align*}
			and if $z \downarrow z_{p_z+1}^+$, then
			\begin{align*}
			-\frac{\left(\sum_{i=1}^{p_z + 1} \eta_i\right) - 1 - z}{p_z + 1} \downarrow \eta_{p_z + 1}.
			\end{align*}
			A similar continuity holds for the second term of (\ref{der0}) due to \eqref{an+1}. The derivative (\ref{der0}) is increasing in $z$, but it is negative for $0<z<z_m$ due to \eqref{boundp} and \eqref{boundn}, which imply
			\begin{align*}
			\frac{\rd Q\left(\vbeta^{(p_z,n_z,z)}\right)}{\rd z}&\leq -2\eta_{p_z+1}+2\eta_{m-n_z}\leqslant 0.
			\end{align*}
			We summarize these results in a proposition.
			\begin{prp}
				The function $Q(\vbeta^{(p_z, n_z, z)})$ is continuous in $z$ for $0 \leq z \leq z_m$, and the derivative with respect to $z$ is negative for $0 < z < z_m$. \label{propref}
			\end{prp}
			
			As $Q(\vbeta^{(p_z,n_z,z)})$ is strictly decreasing in $z$ over $0 < z \leq z_m$ if $p_z + n_z \leq k$, we conclude that $\vbeta^{(p_s,n_s,s)} \in \argmin_{\vbeta\in\calT}Q(\vbeta)$ if $p_s+n_s\leqslant k$. 
			
		\subsection{The case that $p_s + n_s > k$} \label{sssec:solutions}
			
			If $p_s + n_s > k$, then $\vbeta^{(p_s, n_s, s)} \not\in \calT$. So an alternative approach is required. By Lemma \ref{lem1} and the fact that $p_s + n_s > k$, we should compare the objective values for all pairs $(p, n)$ for which $p + n = k$, $p \leqslant p_s$ and $n \leqslant n_s$. In order to do so for a given pair $(p, n)$, we need to find the value of $z$ that minimizes $Q(\vbeta^{(p,n,z)})$. This minimizing value, which we will denote by $\tilde{z}$, must satisfy $z_{pn}^* := \max\{z_p^+, z_{m-n+1}^-\} < \tilde{z} \leqslant s$. We will now show that $\tilde{z}$ is either equal to $s$ or to $s_{pn} := \argmin_z Q(\vbeta^{(p, n, z)})$. \\
			
			We find
			\begin{align*}
			Q(\vbeta^{(p, n, z)}) = (p+n)\left(\frac{\sum_{i=1}^{p} \eta_i + \sum_{j=1}^n \eta_{m-j+1} - 1}{p+n}\right)^2 + \frac{p+n}{pn} \left(s_{pn} - z\right)^2 + \sum_{i=p+1}^{m-n} \eta_i,
			\end{align*}
			where
			\begin{align*}
			s_{pn} = n\frac{\left(\sum_{i=1}^{p} \eta_i\right) - 1}{p+n} - p\frac{\sum_{i=1}^{n} \eta_{m-i+1}}{p+n}.
			\end{align*}
			
			As $Q(\vbeta^{(p, n, z)})$ is quadratic in $z$ with a minimum at $s_{pn}$, we find that if $z^*_{pn}<s\leqslant s_{pn}$, then $\tilde z=s$, and if $z^*_{pn}<s_{pn}<s$, then $\tilde z=s_{pn}$. \\
			
			In the case that $s_{pn}\leqslant z^*_{pn}$, the minimum does not exist, since $Q(\vbeta^{(p, n, z)}) \downarrow Q(\vbeta^{(p, n, z_{pn}^*)})$ as $z \downarrow z_{pn}^*$. Furthermore, $\|\vbeta^{(p, n, z_{pn}^*)}\|_0 < k$. So if $p_{z_{pn}^*} + n_{z_{pn}^*} < k$ then $Q(\vbeta^{(p, n, z_{pn}^*)}) \geq Q(\vbeta^{(p_{z_{pn}^*}, n_{z_{pn}^*}, z_{pn}^*)}) > Q(\vbeta^{(p_z, n_z, z)})$ for some $z > z_{pn}^*$, by Proposition \ref{prp3} and due to the negative gradient of $Q(\vbeta^{(p_z, n_z, z)})$. In the case that $p_{z_{pn}^*} + n_{z_{pn}^*} \geq k$, then $z_{pn}^* = z_p^+$ or $z_{pn}^* = z_n^-$, and $z_p^+ \neq z_n^-$. So if $z_{pn}^* = z_p^+$, then $Q(\vbeta^{(p, n, z_{pn}^*)}) = Q(\vbeta^{(p - 1, n, z_{pn}^*)}) > Q(\vbeta^{(p - 1, n + 1, z_{pn}^*)})$ by Lemma \ref{lem1}. Similarly if $z_{pn}^* = z_n^-$, then $Q(\vbeta^{(p, n, z_{pn}^*)}) > Q(\vbeta^{(p + 1, n - 1, z_{pn}^*)})$. So if $s_{pn} \leqslant z_{pn}^*$, then $\vbeta^{(p, n, z)} \notin \argmin_{\vbeta\in\calT} Q(\vbeta)$ for all $z_{pn}^* < z \leqslant s$. \\
			
			Hence, if $p_s + n_s > k$, we can compute $\tilde{z}$ for each pair $(p, n)$ that satisfies $p + n = k$, $p \leqslant p_s$ and $n \leqslant n_s$ and use this to compute the objective value $Q(\vbeta^{(p, n, \tilde{z})})$. By comparing the objective values, we can find the triplet $(p, n, \tilde{z})$ for which $\vbeta^{(p, n, \tilde{z})} \in \argmin_{\vbeta \in \calT} Q(\vbeta)$. \\
			
			Combining these findings with the findings from the previous sections, we can construct an algorithm to find an element of $\argmin_{\vbeta \in \calT} Q(\vbeta)$. This algorithm is presented in Algorithm \ref{alg:2}. \\
			
		\begin{algorithm}[H]
			\KwIn{Sorted $m$-vector $\veta$, parameters $k$ and $s$.}
			\KwOut{$\widehat{\vbeta}$.}
			$\bar{p} = \argmax_i(i\ |\ \sum_{j = 1}^{i - 1} (\eta_j - \eta_i) < 1 + s)$  \\
			$\bar{n} = \argmax_i(i\ |\ \sum_{j=1}^{i-1} (\eta_{m-i+1}-\eta_{m-j+1}) < s)$ \\
			\uIf{$\bar{p} + \bar{n} \leq k$}{
				$\widehat{\vbeta} = \vbeta^{(\bar{p}, \bar{n}, s)}$
			}
			\Else {
				$\calS = \{(p, n)\ |\ p + n = k,\ p \leq \bar{p},\ n \leq \bar{n}\}$\\
				\For{$(p, n) \in \calS$}{
					$s_{pn} = n\frac{\left(\sum_{i=1}^{p} \eta_i\right) - 1}{p+n} - p\frac{\sum_{i=1}^{n} \eta_{m-i+1}}{p+n}$ \\
					\uIf{$s < s_{pn}$}{
						$z_{pn} = s$	
					}
					\Else{
						$z_{pn} = s_{pn}$
					}
					$Q_{pn} = \|\veta - \vbeta^{(p, n, z_{pn})}\|_2^2$ \\
				}
				$(\widehat{p}, \widehat{n}) = \argmin_{(p, n) \in \calS} Q_{pn}$ \\
				$\widehat{\vbeta} = \vbeta^{(\widehat{p}, \widehat{n}, z_{\widehat{p}\widehat{n}})}$
			}
			\caption{Computing an element of $\argmin_{\vbeta \in \calT} \|\veta - \vbeta\|_2^2$}
			\label{alg:2}
		\end{algorithm}

			\vspace{0.5cm}
		\subsection{Extension}\label{toegift}
			The case $k \leqslant m$ can be treated in a way similar to the case $k<m$, except that in the proof of Proposition \ref{prp2} the assumption $k<m$ was needed. We therefore provide a proof for $k=m$.
			\begin{prp}\label{prp4}
				Proposition \ref{prp2} holds true when $k=m$.
			\end{prp}
			The proof is given in the Appendix.

		\section{Sparsity Under Orthogonality}\label{sec:sparsity}
			In this section, we use the results from Section \ref{compH} to study the sparsity of the solution to \eqref{problem} under orthogonality. \\
			
			As both $\ell_0$- and $\ell_1$-regularization produce sparsity, we can analyze how the sparsity of the solution to \eqref{problem} depends on the tuning parameters $k$ and $s$. From Algorithm \ref{alg:2}, it is straightforward to observe that the amount of non-zero elements in $\widehat{\vbeta}$ is equal to $\min(k, \bar{p} + \bar{n}),$ where $\bar{p} = \argmax_i(i\ |\ \sum_{j = 1}^{i - 1} (\eta_j - \eta_i) < 1 + s)$ and $\bar{n} = \argmax_i(i\ |\ \sum_{j=1}^{i-1} (\eta_{m-i+1}-\eta_{m-j+1}) < s)$.  So the $\ell_1$-regularization component only produces additional sparsity if $k > \bar{p} + \bar{n}$. \\
			
			In order to gain some insights into the sparsity produced by the $\ell_1$-regularization component, we consider the maximum sparsity produced by $\ell_1$-regularization if $k \geq \bar{p} + \bar{n}$. Notice that the sparsity is maximized if $\bar{p} + \bar{n}$ is minimized, which happens when $s = 0$. Furthermore, if $s = 0$, then $\bar{n} = 0$. So, the minimum number of non-zero elements is equal to
			\begin{align}
				\min(\bar{p}, k) 
					&= \bar{p} \nonumber\\
					&= \argmax_i\left(i\ \middle|\ \sum_{j=1}^{i-1}(\eta_j - \eta_i) < 1\right) \nonumber\\
					&= \argmax_i\left(i\ \middle|\ \sum_{j=1}^{i-1}j\Delta_j < 1\right), \label{min:nz}
			\end{align}
			where $\Delta_j = \eta_j - \eta_{j+1}$. This shows that the maximum sparsity produced by $\ell_1$-regularization depends entirely on the size of the gaps between the $\bar{p} + 1$ largest elements of $\veta$. So the maximum amount of sparsity does not change if the same constant is added to each element of $\veta$.\\
			
			To further analyze the maximum sparsity produced by the $\ell_1$-regularization component, we consider two special cases of $\veta$: one case without noise and one case with noise. \\
			
			\textbf{Linear and Noiseless.}
			Suppose that the $\bar{p} +1$ largest elements of $\veta$ are linearly spaced with distance $\Delta > 0$  (i.e. $\eta_i = a - (i - 1)\Delta$ for some $a$). Then, using \eqref{min:nz}, we can derive the following closed-form expression for the minimum number of non-zero elements:
			\begin{align*}
				\bar{p} = \floor[\Bigg]{\frac{1}{2} \left(\sqrt{\frac{\Delta + 8}{\Delta}} + 1\right)},
			\end{align*}
			where $\floor{\cdot}$ rounds down to the nearest integer. As this function is weakly decreasing in $\Delta$, the maximum sparsity is increasing in $\Delta$. So, we obtain the intuition that if the largest elements of $\veta$ are more similar, then less sparsity can be produced by $\ell_1$-regularization. \\
			
			\textbf{Equal and Noisy.}
			Let $\veta = \vbeta^*+ \sigma\vepsi$, where $\vepsi$ has i.i.d. elements $\varepsilon_i \sim N(0, 1)$, $\sigma > 0$, and $\vbeta^*$ is an $m$-vector with elements $\beta_i^* = \beta_j^*$ for all $i, j$. As all elements of $\vbeta^*$ are equal, the gaps between the elements of $\veta$ are equal to the gaps between the order statistics of $\vepsi$, scaled by the constant $\sigma$. So, the size of the gaps between the largest elements of $\veta$ is increasing in $\sigma$. Therefore, according to \eqref{min:nz}, the maximum sparsity is increasing in $\sigma$. As an increase in $\sigma$ represents an increase in noise, we can draw the intuitive conclusion that if $\vbeta^*$ has elements of similar size, then the maximum amount of sparsity produced by $\ell_1$-regularization increases with noise.
			
			
	\section{General Case}
		\label{sec:DFO}
		In this section, we describe how a solution can be found for the general case, in which $\mX$ is not required to be orthogonal. To do so, we adapt the framework laid out by \citet{bertsimas2016best} for standard linear regression. This framework consists of two components. The first component is a Discrete First-Order (DFO) algorithm that uses an algorithm for the orthogonal problem as a subroutine in each iteration. The solution to this DFO algorithm is then used as an initial solution for the second component. The second component relies on reformulating (\ref{problem}) as an MIO problem, which can be solved to provable optimality by using an MIO solver.
		
		\subsection{Discrete First-Order Algorithm}
			In the construction of the DFO algorithm, we closely follow \citet{bertsimas2016best}, but use a different constraint set that includes an additional $\ell_1$-norm restriction and unit-sum restriction. \\
			
			Denote the objective function as
			\begin{align*}
				f(\vbeta) = \frac{1}{2}\|\vy - \mX\vbeta\|_2^2.
			\end{align*}
			
			This function is Lipschitz continuously differentiable, as
			\begin{align*}
				\|\nabla f(\vbeta) - \nabla f(\veta)\|_2^2 
				&= \|\mX'\mX (\vbeta - \veta) \|_2^2 \\
				&\leq \|\mX'\mX\|^2 \|\vbeta - \veta \|_2^2 \\
				&= L^*\|\vbeta - \veta \|_2^2,
			\end{align*}
			where $L^*$ is the largest absolute eigenvalue of $\mX'\mX$. So, we can apply the following result.
			\begin{prp}[\citealp{nesterov2013introductory, bertsimas2016best}]\label{prp:bert}
			 For a convex Lipschitz continuous function $f(\cdot)$, we have 
				\begin{align}
					f(\veta) \leq Q_L(\veta, \vbeta) := f(\vbeta) + \frac{L}{2} \|\veta - \vbeta\|_2^2 + \nabla f(\vbeta)' (\veta - \vbeta),
					\label{bound}
				\end{align}
				
				for all $L \geq \bar{L}$, $\vbeta$ and $\veta$, where $\bar{L}$ is smallest constant such that $\|\nabla f(\vbeta) - \nabla f(\veta)\|_2^2 \leq  \bar{L}\|\vbeta - \veta \|_2^2$.
			\end{prp}
			\vspace{0.5cm}
			Given some fixed $\vbeta$, we can minimize the bound in (\ref{bound}) with respect to $\veta$ under the constraint set $\calT$, as given in \eqref{def:T}. Following \citet{bertsimas2016best}, we find
			\begin{align}
				\hspace{-1cm}
				\argmin_{\veta \in \calT} Q_L(\veta, \vbeta)
					&= \argmin_{\veta \in \calT} \left(f(\vbeta) + \frac{L}{2} \|\veta - \vbeta\|_2^2 + \nabla f(\vbeta)' (\veta - \vbeta) + \frac{1}{2L} \|\nabla f(\vbeta)\|_2^2 - \frac{1}{2L} \|\nabla f(\vbeta)\|_2^2 \right) \nonumber\\
					&= \argmin_{\veta \in \calT} \left(f(\vbeta) + \frac{L}{2} \|\veta - (\vbeta - \frac{1}{L} \nabla f(\vbeta))\|_2^2 - \frac{1}{2L} \|\nabla f(\vbeta)\|_2^2 \right) \nonumber\\
					&= \argmin_{\veta \in \calT} \|\veta - (\vbeta - \frac{1}{L} \nabla f(\vbeta))\|_2^2. \label{finalterm}
			\end{align}
			Notice that \eqref{finalterm} can be computed using Algorithm \ref{alg:2}. Therefore, it is possible to use iterative updates in order to decrease the objective value. Specifically, let $\vbeta_1 \in  \calT$ and recursively define $\vbeta_{r+1} = \argmin_{\veta \in \calT} Q_{L^*}(\veta, \vbeta_r)$, for all $r \in \mathbb{N}$. Then by Proposition \ref{prp:bert},
			\begin{align*}
				f(\vbeta_{r}) 
					= Q_{L^*}(\vbeta_r, \vbeta_r)
					\geq  Q_{L^*}(\vbeta_{r+1}, \vbeta_r)
					\geq f(\vbeta_{r+1}).
			\end{align*}
			In Algorithm \ref{alg:1}, we present an algorithm that uses this updating step until some convergence criterion is reached. \\
			
			\begin{algorithm}[H]
				\KwIn{Lipschitz constant $L^*$, convergence criterion $\varepsilon$, initial solution $\vbeta_1 \in \calT$.}
				\KwOut{$\widehat{\vbeta}$}
				$r = 1$ \\
				\Repeat{$f(\vbeta_{r}) - f(\vbeta_{r-1}) < \varepsilon$}{
					$\vbeta_{r+1} \in \argmin_{\veta \in \calT} \|\veta - (\vbeta_r - \frac{1}{L^*} \nabla f(\vbeta_r))\|_2^2$ (using Algorithm \ref{alg:2}). \\
					$r = r + 1$
				}
				$\widehat{\vbeta} = \vbeta_r$
				\caption{First order algorithm\vspace{.1cm}}
				\label{alg:1}
			\end{algorithm}
			\vspace{0.5cm}

		\subsection{Mixed-Integer Optimization}
			\label{sec:MIO}
			
			In this section, an MIO formulation for problem \eqref{problem} is presented. In order to formulate problem (\ref{problem}) as an MIO problem, we use three sets of auxiliary variables. The variables $\beta_i^+$ and $\beta_i^-$ are used to specify the positive and negative parts of the arguments $\beta_i$, $i \in \{1, \dots, m\}$. The variable $z_i$ serves as an indicator function for whether $\beta_i$ is different from zero, $i \in \{1, \dots, m\}$. The MIO formulation is given as follows:
			\begin{align*}
				\min_{\vbeta, \vz} &\ \vbeta'\mX'\mX\vbeta - 2 \vy'\mX\vbeta + \lambda \vbeta'\vbeta, \\
				\quad\text{s.t.} \\
					&\beta_i = \beta_i^+ - \beta_i^-, \quad i \in \{1, \dots, m\}, \\
					&\sum_{i=1}^m \beta_i = 1, \\
					&\sum_{i=1}^m \beta_i^+ \leq 1 + s, \\
					&\sum_{i=1}^m \beta_i^- \leq s, \\
					&\calM_-z_i \leq \beta_i \leq \calM_+z_i, \quad i \in \{1, \dots, m\}, \\
					&\sum_{i=1}^m z_i \leq k, \\
					&\beta_i^+, \beta_i^- \geq 0,\quad i \in \{1, \dots, m\}, \\
					&z_i \in \{0, 1\},\quad i \in \{1, \dots, m\},
			\end{align*}
			where $\vbeta$ has elements $\beta_i$, and $\calM_+$ and $\calM_-$ are big-M parameters. These big-M parameters are used to enforce the sparsity constraint as follows: if $z_i = 1$ then $\beta_i \in [\calM_-, \calM_+]$, and if $z_i = 0$ then $\beta_i = 0$. Hence, $\calM_-$ and $\calM_+$ should be sufficiently large in absolute value to ensure that the solution to the MIO problem is the solution to \eqref{problem}. On the other hand, they should not be too large as tighter bounds decrease the size of the search space and improve the speed of the solver. \\
			
			The $\ell_1$-restriction provides natural choices $\calM_+ = 1 + s$ and $\calM_- = -s$. However, these bounds are conservative in practice. \citet{mazumder2017subset} suggest the use of bounds based on the solution of the DFO algorithm. Similarly, we propose to use $\calM_- = \max\{\frac{3}{2} \min(\min_i[\beta_i^{\text{DFO}}], -\frac{s}{k}), -s\}$ and $\calM_+ = \min\{\frac{3}{2} \max(\max_i[\beta_i^{\text{DFO}}], \frac{1 + s}{k}), 1 + s\}$, where $\beta_i^{\text{DFO}}$ is the $i$th element of the solution of the DFO algorithm.

	\section{Numerical Results}\label{sec:num}
		In this section we compare the performance of our $\ell_0\ell_1$-regularized approach to $\ell_0$-regularization and $\ell_1$-regularization on simulated datasets, generated with multiple signal-to-noise ratios and values of $\vbeta$.
		
		\subsection{Setup Simulation Experiments}\label{subsec:sim}
			The setup of our simulation experiments largely follows the numerical experiments found in \citet{mazumder2017subset} and \citet{hastie2017extended}. For a given set of parameters $t$ (number of observations), $m$ (number of variables), $k^*$ (number of non-zero weights), $p$ (number of positive weights), $n$ (number of negative weights), $s^*$ (sum of the negative weights), $\rho$ (autocorrelation between the variables) and SNR (signal-to-noise ratio), the experiments are conducted as follows:
			
			\begin{enumerate}
				\item[1.] We randomly select $k^*$ elements of $\vbeta$ and set $p$ of the elements equal to $(1 + s^*)/p$, and $n$ of the elements equal to $-s^*/n$. The remaining elements are set equal to zero.
				\item[2.] The rows of $t \times m$ matrix $\mX$ are drawn i.i.d. from $N_m(\vzeros, \mSigma)$, where $\mSigma$ has elements $\sigma_{ij} = \rho^{|i - j|}$, $i, j \in \{1, \dots, p\}$.
				\item[3.] The vector $\vy$ is drawn from $N(\mX\vbeta, \sigma^2\mI)$, where $\sigma^2 = \vbeta'\mSigma\vbeta / \text{SNR}$ in order to fix the signal-to-noise ratio.
				\item[4.] We apply $\ell_0$-regularization, $\ell_1$-regularization and $\ell_0\ell_1$-regularization to $\mX$ and $\vy$ for a range of tuning parameters. For both methods, we select the tuning parameter(s) that minimize(s) the prediction error on a separate dataset $\widetilde{\mX}$, $\widetilde{\vy}$, generated in the same way as $\mX$ and $\vy$. 
				\item[5.] We record several performance measures of the solutions that were found using the selected tuning parameters.
			\end{enumerate}
			
			We repeat the above steps 100 times for each parameter setting. Throughout the experiments we use $t = 50$, $m = 100$, $k^* = 7$, $\rho = 0.2$. For each setting, we choose $s^* \in \{0.1, 2/3\}$ and SNR $\in \{2^{-1}, 2^{0}, 2^{1}\}$. This choice of $s^*$ covers the case where the negative weights are small in comparison to the positive weights, as well as the case where the positive and negative weights are equal in magnitude. The tuning parameters corresponding to $k^*$ and $s^*$ are simultaneously selected over the grid $\{1, \dots, 20\} \times \{0, s^* / 5, \dots, 2s^*\}$.  \\
			
			For each different combination of $s^*$ and SNR, we record the following performance measures:
			\begin{enumerate}
				\item[-] \textbf{Relative risk}. As measure of predictive performance we use relative risk, defined for a solution $\widehat{\vbeta}$ as
				\begin{align*}
					\text{RR}(\vbeta) = \frac{(\widehat{\vbeta} - \vbeta)'\mSigma(\widehat{\vbeta} - \vbeta)}{\vbeta\mSigma\vbeta}.
				\end{align*}
				This is one of the measures used in \citet{hastie2017extended}, and is similar to the predictive performance measures used in \citet{bertsimas2016best} and \citet{mazumder2017subset}. For this measure, a lower value is indicative of a better predictive performance and its minimum value is 0. The null score to beat is 1 (if $\widehat{\vbeta} = \vzeros$).
				\item[-] \textbf{Number of non-zero elements}. As a second measure, we consider the number of non-zero elements in the estimated weights, in order to compare the sparsity obtained by both methods.
				\item[-] \textbf{Sum of negative weights}. As a final measure, we consider the sum of the negative estimated weights. This allows us to compare the shrinkage produced by the $\ell_1$-regularization component of both methods.
			\end{enumerate}

		\subsection{Implementation and Stopping Criteria} \label{subsec:initialization}
			In order to compute the solution to $\ell_1$-regularized unit-sum regression, we use an adaptation of the LARS method for $\ell_1$-regularization \citep{efron2004least}, based on the algorithm described by \citet{demiguel2009generalized}. The $\ell_0$-regularization solution is computed in the same way as the $\ell_0\ell_1$-regularization solution by fixing the parameter $s$ to some sufficiently large value. \\
			
			For the $\ell_0\ell_1$-regularization approach, we terminate the DFO algorithm if the improvement in the squared error is below some value $\varepsilon$, where we set $\varepsilon = 10^{-6}$. As the DFO algorithm can be sensitive to its initialization, we initialize it with the Forward-Stepwise Selection (FSS). We found that this typically yields a better performance than using the best solution out of 50 random initializations as used by \citet{bertsimas2016best}. The FSS solution is implemented using successive applications of the adapted LARS algorithm. \\
			
			The MIO formulation is implemented in the R-interface of Gurobi 7.1. Each instance is given 10 minutes of computation time. If the optimality of the solution is not confirmed within the allotted time, the solver is terminated and its best solution so far is used. This means that the combined maximum computation time is 44000 hours. However, in practice we find that the DFO algorithm often provides optimal or near-optimal solutions to the MIO solver. As a result, the MIO solver rarely uses the full 10 minutes and typically certifies optimality in seconds. The total computation time for the simulation experiments was approximately 600 hours on a single machine, including the computation of the initial solutions.
		
		\subsection{Results of Simulation Experiments}\label{sec:res}

\begin{figure}
		\centering
		Setting $s^* = 0.1$\\
		\includegraphics[width=\linewidth]{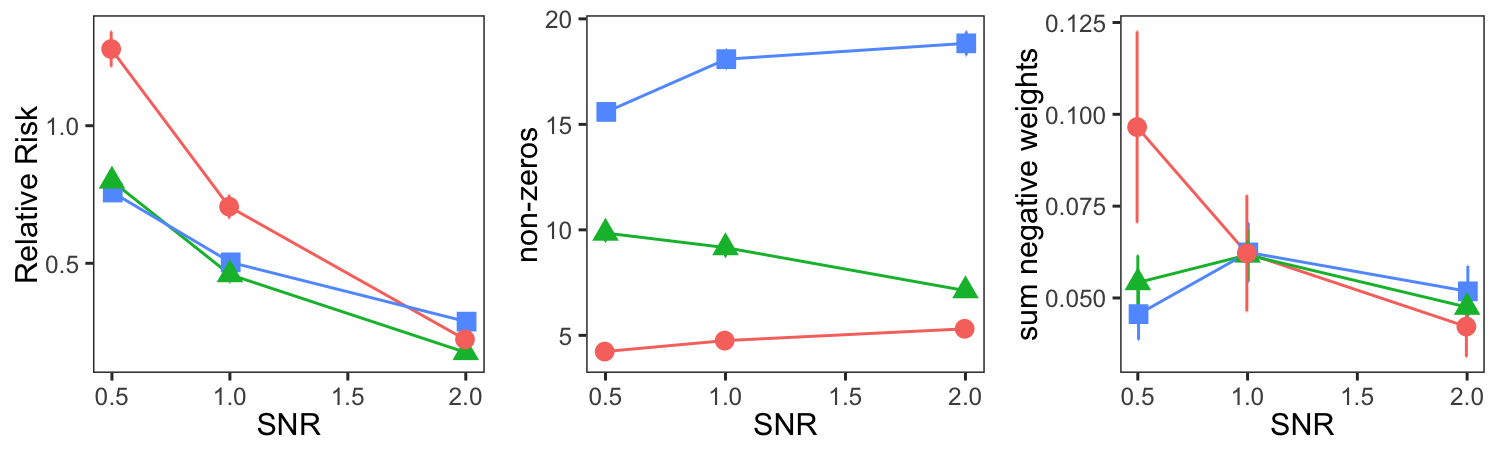}
		Setting $s^* = 2/3$\\
		\includegraphics[width=\linewidth]{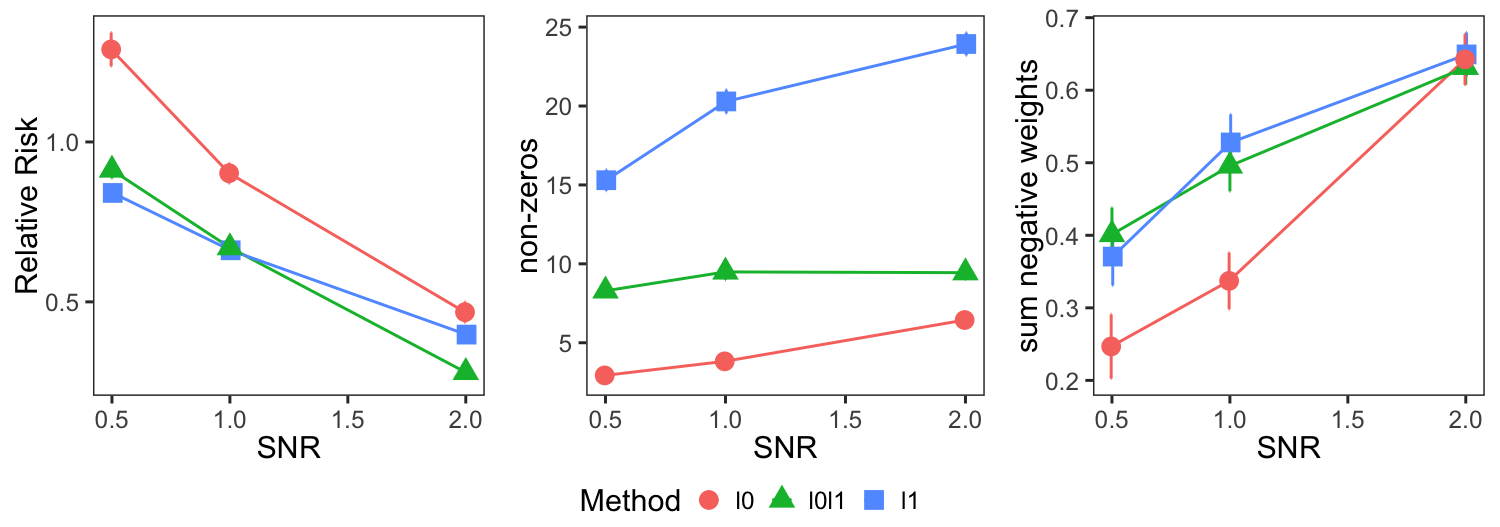}
		\caption{Performance of $\ell_0$-regularization, $\ell_1$-regularization and $\ell_0\ell_1$-regularization in terms of relative risk, number of non-zero elements and the sum of the negative weights as a function of the SNR for two values of $s^*$. The vertical bars represent one standard error.}
		\label{plot:1}
	\end{figure}

			The results of the simulation experiments are displayed in Figure \ref{plot:1}. We make the following observations. \\
			
			\textbf{Prediction}.
			It can be observed that $\ell_0$-regularization typically performs worse than the other methods, especially when the SNR is low. Furthermore, $\ell_0\ell_1$-regularization seems to outperform $\ell_1$-regularization for higher SNRs in terms of relative risk, while $\ell_1$-regularization fares similarly or even somewhat better for lower SNRs.\footnote{These results differ slightly from the findings by \citet{mazumder2017subset} for standard linear regression. They find that $\ell_0\ell_1$-regularization performs as well as $\ell_1$-regularization if the SNR is low. We suspect that this difference could be caused by the fact that they do not consider an SNR below 1 and use a fixed-design setup where $\widetilde{\mX} = \mX$.} \\
			
			\textbf{Sparsity.} 
			We find that $\ell_1$-regularization delivers considerably denser solutions than the other methods for all values of SNR and $s^*$. In addition, the number of non-zeros seems to move away from the true number of non-zeros as the SNR increases. On the other hand, $\ell_0$-regularization yields overly sparse solutions below the true value $k^*$, especially if the SNR is low. The number of non-zeros produced by $\ell_0\ell_1$-regularization lies between the values other two methods, and is typically closer $k^*$ than the number of non-zeros produced by $\ell_0$-regularization or $\ell_1$-regularization. \\
			
			\textbf{Shrinkage.} 
			In the third column of Figure \ref{plot:1}, it can be seen that the sum of the negative weights of the solutions tends to be smaller than $s^*$.  However, for the case that $s^* = 2 / 3$, there is a clear trend towards the true value of $s^*$ as the SNR increases. Interestingly, both $\ell_0\ell_1$-regularization and $\ell_1$-regularization have a similar sum of negative weights, despite the fact that the solutions of $\ell_0\ell_1$-regularization are much sparser. This implies that the average magnitude of the weights of the $\ell_1$-regularization solution is much smaller than that of the $\ell_0\ell_1$-regularization solution. 
		
	\section{Application: Index Tracking}\label{subsec:maxsparse}
		In order to demonstrate the use of our proposed methodology in practice, we consider an application to index tracking. Index tracking concerns the construction of a portfolio that replicates a stock index as closely as possible, while limiting the cost of holding the portfolio. Such a portfolio can be represented by a weight vector that sums to one, with positive elements that correspond to long positions and negative elements that correspond to short positions. \\
		
		Two standard ways to limit the cost of holding the portfolio are to restrict the number of constituents in the portfolio and to avoid short positions. Using historical returns data, it is possible to find such a portfolio using $\ell_0\ell_1$-regularized unit-sum regression. Specifically, let $\vy$ represent the historical returns of a stock index and let each column of $\mX$ represent the historical returns of one of its constituents. Then, using $s=0$, problem \eqref{problem} minimizes the squared error between the actual index returns and the returns of the portfolio, that consists of at most $k$ constituents and has no short positions. \\

		Notice that even if the $\ell_0$ component is omitted, or equivalently $k = m$, then the remaining $\ell_1$-regularization may still produce a sparse portfolio \citep{demiguel2009generalized, brodie2009sparse}. However, as the returns of an index are typically a dense linear combination of its constituent returns, with positive weights of similar magnitude, the intuitions from the orthogonal design case from Section \ref{sec:sparsity} suggest that $\ell_1$-regularization may not be very effective in producing sparsity. \\

		To compare the sparsity production and tracking performance of $\ell_0\ell_1$-regularization and $\ell_1$-regularization, we use the index tracking datasets of the OR-library \citep{beasley2003evolutionary, canakgoz2009mixed}. These datasets contain 290 weekly returns of 8 indexes varying from 31 to 2153 constituents.\footnote{Only the constituents that are part of the index for the entire period are included.} Each dataset is split into two halves of 145 observations, where the first half is used to construct the portfolio and the second half is used to measure the performance of the portfolio.\footnote{From the second index (DAX) we removed two large consecutive outliers from the out-of-sample data. These two outliers were the largest two returns (in absolute value) and of opposing sign, suggesting a bookkeeping error in the index returns. This is supported by the fact that the outliers are not reflected in the returns of the constituents.} The performance is measured in out-of-sample $R^2$, on the second half of the datasets. The results are presented in Table \ref{tab:benchmarks}. \\
		
		From the results we can make several observations regarding the sparsity of the solutions and the tracking performance. First it should be noted that $\ell_1$-regularization by itself is not able to find a unique portfolio for the largest two stock indexes. In addition, even if $\ell_1$-regularization does have a unique solution, it is generally not able to produce a substantial amount of sparsity. In terms of out-of-sample tracking performance, lower values of $k$ do generally result in worse performance. However, the difference is small, especially for the larger values of $k$. \\

	\clearpage
	\section{Appendix}\label{Appendix}
		\appendix

		\setcounter{equation}{0}
		\def\theequation{A.\arabic{equation}}
		\setcounter{section}{0}
		\def\thesection{A}
		
		\textbf{Proof of Proposition \ref{prp2}}:
		If $\vbeta^{(\calP, \calN, z)} \in \argmin_{\vbeta \in \calA_z} Q(\vbeta)$, then $\calP = \{1, \dots, p\}$ and $\calN = \{m-n+1, \dots, m\}$ if $n\geqslant 1$. 		
		\begin{proof}
			We show that if $\vbeta^{(\calP, \calN, z)} \in \argmin_{\vbeta \in \calA_z} Q(\vbeta)$, then two conditions hold true:
			\begin{align*}
			&\mbox{$\calP$-condition}:\quad \max(\calP)<\min\{(\calP\cup\calN)^C\},\\
			&\mbox{$\calN$-condition}:\quad \min(\calN)>\max\{(\calP\cup\calN)^C\}\ \ \mbox{if}\ \ n\geqslant 1.
			\end{align*}
			We prove the $\calP$-condition. The proof of the $\calN$-condition is similar. Assume the $\calP$-condition is not true. In that case, we show that an index set $\overline{\calP}$ exists, such that $\vbeta^{(\overline\calP, \calN, z)} \in \calA_z$ and $Q(\vbeta^{(\overline\calP, \calN, z)})<Q(\vbeta^{(\calP, \calN, z)})$, which is a contradiction, showing the validity of the $\calP$-condition.
			
			\hspace{1cm}Assuming the $\calP$-condition is not true, let $u:=\max(\calP)>v:=\min\{(\calP\cup\calN)^C\}$. As $\vbeta^{(\calP, \calN, z)} \in \calA_z$, we have $\beta_u^{(\calP,\calN,z)}>0$,
			which is equivalent to $z+1>\sum_{i\in\calP}(\eta_i-\eta_u)$.
			Define 
			\begin{align*}
			\calP^*&:=\calP\setminus u,\\
			\widetilde\calP_j&:=\{i\ |\ i\in\calP^*,\ i\leqslant j\in\calP^*\},\\
			\overline \calP_j&:=\widetilde\calP_j\cup v.
			\end{align*}
			Let $j_v:=\max\{j\ |\ j\in\calP^*,\ j<v\}$. As $\eta_i-\eta_u> 0$ if $i\in\calP^*$, we find
			\begin{align*}
			z+1>\sum_{i\in\calP}(\eta_i-\eta_u)=\sum_{i\in\calP^*}(\eta_i-\eta_u)\geqslant\sum_{i\in\widetilde{\calP}_{j_v}}(\eta_i-\eta_u)>\sum_{i\in\widetilde{\calP}_{j_v}}(\eta_i-\eta_v)=\sum_{i\in\overline\calP_{jv}}(\eta_i-\eta_v).
			\end{align*}
			Consequently, $\vx^{(\overline{\calP}_{j_v},\calN,z)}\in\calA_z$. Therefore, the index set $\overline\calP_{\bar j}$, where $\bar j$ is the maximum index such that $\vx^{(\overline{\calP_{\bar j}},\calN,z)}\in\calA_z$, exists. Let $\overline\calP=\overline\calP_{\bar j}$.
			
			\hspace{1cm}We now show $Q(\vbeta^{(\overline\calP, \calN, z)})<Q(\vbeta^{(\calP, \calN, z)})$. Let $\calR=\calP^*\setminus\widetilde\calP_{\bar j}$ with cardinality $r$. As $\calP^*=\widetilde{\calP}_{\bar j}\cup\calR$ has cardinality $p-1$, the cardinalty of $\overline\calP=\widetilde{\calP}_{\bar j}\cup v$ equals $p-r$. Let
			\begin{align*}
			b&:=\frac{\left(\sum_{i\in\calP}\eta_i\right) -1-z}{p},\quad\mbox{and}\quad
			\bar{b}:=\frac{\left(\sum_{i\in\overline\calP}\eta_i\right) -1-z}{p-r}.
			\end{align*}
			We find
			\begin{align*}
			Q(\vbeta^{(\overline\calP, \calN, z)})-Q(\vbeta^{(\calP, \calN, z)})&= (p-r)\bar b^2 - p b^2 + \eta_u^2 - \eta_v^2 + \sum_{i\in\calR}\eta_i^2 \\								
			&= (p-r)\left(\bar b + b\right)\left(\bar b - b\right) 
			+ \eta_u^2 - \eta_v^2   + \sum_{i\in\calR}(\eta_i^2-b^2)\\
			&= \left(\bar b + b\right)\left\{\eta_v - \eta_u - \sum_{i\in\calR}(\eta_i - b)\right\}
			+ \eta_u^2 - \eta_v^2  +  \sum_{i\in\calR}(\eta_i^2-b^2) \\																	
			&=\left(\eta_u + \eta_v - \bar b - b\right)\left(\eta_u - \eta_v\right) 
			+ \sum_{i\in\calR}(\eta_i-b)(\eta_i-\bar b) \\								
			&= \left(\beta_u^{(\calP, \calN, z)} + \beta_v^{(\overline\calP, \calN, z)}\right)(\eta_u - \eta_v) + \sum_{i \in \calR} \beta_i^{(\calP, \calN, z)} \left(\eta_i - \bar b\right). 									
			\end{align*}
			As $\beta_u^{(\calP, \calN, z)}>0$, $\beta_v^{(\overline\calP, \calN, z)}>0$ and $\eta_u<\eta_v$, the first term is negative. As $\beta_i^{(\calP, \calN, z)}>0$ and $\eta_i - \bar b<0$ if $i\in\calR$, and hence $i>\bar{j}$, the second term is negative as well. Consequently, $Q(\vbeta^{(\overline\calP, \calN, z)})<Q(\vbeta^{(\calP, \calN, z)})$, which is a contradiction.
		\end{proof}
		
		\textbf{Proof of Lemma \ref{lem1}}: If $\vbeta^{(\tilde{p}, \tilde{n}, z)} \in \calA_z$ and $\vbeta^{(p, n, z)} \in \calA_z$, where $\tilde{p} \leq p$, $\tilde{n} \leq n$, $\tilde{p} + \tilde{n} < p + n$, then $Q(\vbeta^{(p, n, z)}) < Q(\vbeta^{(\tilde{p}, \tilde{n}, z)})$.
		\begin{proof}
			We can use the convexity of the quadratic function to show
			\begin{align*}
			\hspace{-1cm} Q(\vbeta^{(p,n,z)})-Q(\vbeta^{(p+1,n,z)})
			&= p \left(\frac{\left(\sum_{i=1}^p\eta_i\right)-1-z}{p}\right)^2 + \eta_{p+1}^2 - (p+1) \left(\frac{\left(\sum_{i=1}^{p+1}\eta_i\right)-1-z}{p+1}\right)^2 \\
			&=(p+1) \left[\lambda y_1^2 + (1-\lambda) y_2^2 - \right\{\lambda y_1 + (1 - \lambda) y_2 \left\}^2 \right]>0,
			\end{align*}
			where $\lambda=p/(p+1)$, $y_1=\left\{\left(\sum_{i=1}^p\eta_i\right)-1-z\right\}/p$ and $y_2=\eta_{p+1}$. In a similar way we find $Q(\vbeta^{(p,n,z)})>Q(\vbeta^{(p,n+1,z)})$.
		\end{proof}
		\textbf{Proof of Proposition \ref{prp3}}: If $\vbeta^{(\calP, \calN, z)} \in \argmin_{\vbeta \in \calA_z} Q(\vbeta)$ and $p_z+n_z\leq k$, then $\vbeta^{(\calP, \calN, z)}=\vbeta^{(p_z,n_z,z)}$.	
		
		\textbf{Proof of Proposition \ref{prp4}}: Proposition \ref{prp2} holds true when $k=m$.	
		\begin{proof}
			Using Proposition \ref{prp1}, let $\vbeta^{(\calP,\calN,z)}\in\argmin_{\vbeta\in\calA_z}Q(\vbeta)$ where $\calP$ and $\calN$ have cardinalities $p$ and $n$, respectively, where $p+n=m$. We will show that $u:=\max(\calP)<v:=\min(\calN)$ if $n\geq 1$.\\
			\mbox{}\hspace{1cm}Suppose $u:=\max(\calP)>v:=\min(\calN)$, then
			\begin{align}\label{extra}
				\frac{\left(\sum_{i\in\calP}\eta_i\right)-1-z}{p}<\eta_u< \eta_v<\frac{\left(\sum_{i\in\calN}\eta_i\right)+z}{n}.
			\end{align}
			Consequently,
			\begin{align*} 
				Q\left(\vbeta^{(\calP,\calN,z)}\right)=p\left\{\frac{\left(\sum_{i \in \calP} \eta_i\right)-1-z}{p}\right\}^2+n\left\{\frac{(\sum_{i\in\calN} \eta_{m-i+1})+z}{n}\right\}^2
			\end{align*}
			has a positive derivative
			\begin{align*} 
				\frac{\rd Q\left(\vbeta^{(\calP,\calN,z)}\right)}{\rd z}&=-2\left\{\frac{\left(\sum_{i\in\calP} \eta_i\right)-1-z}{p}\right\}+2\left\{\frac{(\sum_{i\in\calN} \eta_{m-i+1})+z}{n}\right\}\\
			&>2(\eta_v-\eta_u)>0.
			\end{align*}
			As a result, $\{Q\left(\vbeta^{(\calP,\calN,z)}\right)\}\ |\ z\  \mbox{satisfies \eqref{extra}}\}$ does not have a minimum. If 
			\begin{align*}
				z^*=\min\left\{z\ \left|\ \frac{\left(\sum_{i\in\calP}\eta_i\right)-1-z}{p}\leqslant \eta_u<  \eta_v\leqslant\frac{\left(\sum_{i\in\calN}\eta_i\right)+z}{n}\right.\right\},
			\end{align*}
			then $\vbeta^{(\calP^*,\calN^*,z^*)}$ contains zeros, so that $Q(\vbeta^{(\calP^*,\calN^*,z^*)})\geq Q(\vbeta^{(p_{z^*},n_{z^*},z^*)})>Q(\vbeta^{(p_z,n_z,z)})$.
		\end{proof}
	\clearpage
	\clearpage
	\begin{table}[h!]
		\centering
		\begin{tabular}{lrrrclrrr}
			\hline
			\hline
			Index & $k$  &   \multicolumn{1}{c}{\#nz}   &  \multicolumn{1}{c}{$R^2_{\text{oos}}$} &&Index & \multicolumn{1}{c}{$k$}   &   \multicolumn{1}{c}{\#nz}   &  \multicolumn{1}{c}{$R^2_{\text{oos}}$} \\
			\hline
			& 5 & 5 &  0.909 & \quad\quad\quad\quad & & 20 & 20 & 0.922 \\ 
			Hang  & 15 & 15 & 0.982 &  & Nikkei & 60 & 60 & 0.957 \\ 
			Seng  & 25 & 25 & 0.991 &  & (m = 225)& 100 & 100 & 0.961 \\ 
			(m = 31) & 31 & 25 & 0.991&  && 225 & 127 & 0.961 \\ 
			&  &   &   &   &   &   & &   \\ 
			& 10 & 10 & 0.940& && 20 & 20 & 0.780 \\ 
			DAX & 30 & 30 & 0.979 &  &S\&P& 60 & 60 & 0.839 \\ 
			(m = 85) & 50 & 50 & 0.981 &  &500& 100 & 100 & 0.857 \\ 
			& 85 & 78 & 0.985 &  &(m = 457)& 457 & 122 &  0.855 \\ 
			&  &   &   &   &   &   &  & \\ 
			& 10 & 10 & 0.652 &  && 20  & 20 & 0.646 \\ 
			FTSE & 30 & 30 & 0.948 &  &Russel& 60 & 60 & 0.679 \\ 
			(m = 89) & 50 & 50 & 0.959 &  &2000& 100 & 100 & 0.691 \\ 
			& 89 & 68 & 0.966 &  & (m = 1319)& 1319 & - & - \\ 
			&  &   &   &   &   &   &  & \\ 
			& 10 & 10 & 0.815 &  && 20 & 20 & 0.767 \\ 
			S\&P & 30 & 30 & 0.932 &  &Russel& 60& 60 & 0.821 \\ 
			100 & 50 & 50 & 0.960 &  &3000& 100& 100 & 0.836\\ 
			(m = 98) & 98 & 77 & 0.969 &  & (m = 2152)& 2152 & - & - \\
			\hline
			\hline
		\end{tabular}
		\caption{Out-of-sample $R^2$ ($R^2_{\text{oos}}$) and number of non-zeros (\#nz) of $\ell_0\ell_1$-regularized unit-sum regression for all 8 index datasets and multiple values for the parameters $k$, using $s = 0$. The results for $k = m$ are equivalent to $\ell_1$-regularized unit-sum regression with $s = 0$. A hyphen (-) indicates that no unique solution was found.}
		\label{tab:benchmarks}
	\end{table}
	\bibliographystyle{abbrvnat}
	\bibliography{bibfile}	
\end{document}